\documentclass[showpacs,nofootinbib,twocolumn]{revtex4}
\usepackage{graphicx}
\usepackage{amsfonts}
\usepackage{amssymb}
\usepackage{amsmath}

\begin{document}

\preprint{University of Waterloo Preprint}
\title{The Dynamics of Test Particles and Pointlike Gyroscopes \\
in the Brane World and Other 5D Models}
\author{Sanjeev S. Seahra}
\email{ssseahra@uwaterloo.ca}
\affiliation{Department of Physics,
University of Waterloo\\ Waterloo, Ontario, N2L 3G1, Canada}
\date{\today}

\setlength\arraycolsep{2pt}
\newcommand*{\di}{\partial}
\newcommand*{\al}{\alpha}

\renewcommand{\textfraction}{0.15}
\renewcommand{\topfraction}{0.6}

\begin{abstract}

We study the dynamics of test particles and pointlike gyroscopes
in 5D manifolds like those used in the Randall-Sundrum brane
world and non-compact Kaluza-Klein models.  Our analysis is based
on a covariant foliation of the manifold using $3+1$ dimensional
spacetime slices orthogonal to the extra dimension, and is hence
similar to the ADM $3+1$ split in ordinary general relativity. We
derive gauge invariant equations of motion for freely-falling
test particles in the 5D and 4D affine parameterizations and
contrast these results with previous work concerning the
so-called ``fifth force''. Motivated by the conjectured
localization of matter fields on a 3-brane, we derive the form of
the classical non-gravitational force required to confine
particles to a 4D hypersurface and show that the resulting
trajectories are geometrically identical to the spacetime
geodesics of Einstein's theory.  We then discuss the issue of
determining the 5D dynamics of a torque-free spinning body in the
point-dipole approximation, and then perform a covariant
$(3+1)+1$ decomposition of the relevant formulae (i.e.\ the 5D
Fermi-Walker transport equation) for the cases of freely-falling
and hypersurface-confined point gyroscopes.  In both cases, the
4D spin tensor is seen to be subject to an anomalous torque.  We
solve the spin equations for a gyroscope confined to a single
spacetime section in a simple 5D cosmological model and observe a
cosmological variation of the magnitude and orientation of the 4D
spin.

\end{abstract}


\maketitle

\section{Introduction}\label{sec:introduction}

There has been a large amount of recent interest in the
possibility that our world contains more than four non-compact
dimensions.  This interest has been largely motivated by the work
of Randall and Sundrum (RS), who postulate that our 4D universe
is in actuality a 3-brane moving in a higher dimensional manifold
with large extra dimensions \cite{Ran99a,Ran99b}.  The
introduction of such a model assists in the explanation of the
hierarchy and cosmological constant problems.  However, the
existence of large extra dimensions raises several issues, not
the least of which is the nature of the trajectories of test
particles in a $(3+1)+d$ dimensional manifold.\footnote{The
notation is that $(3+1)$ refers to ordinary spacetime with one
timelike and three spacelike dimensions while $d$ refers to the
number of spacelike extra dimensions.}  Several authors have
considered this, both in the context of the $d = 1$ brane world
picture and non-compact 5D Kaluza-Klein theory.  In the former
scenario, matter is confined to the brane while gravitons are
free to propagate in the bulk on 5D null geodesics.  In the
latter scenario, all particles travel on 5D geodesics but
observers can only readily access the 4D part of the trajectory
\cite[reviews]{Wes99a,Wes01a}.  Specific topics tackled in the
literature include deviations from 4D geodesic motion and the
so-called ``fifth force''
\cite{Gro83,Cho91,Mas98,Wes99,You00,You01,Pon01a,Pon01b},
astrophysical and experimental tests of higher dimensional
dynamics \cite{Kal94,Wes97,Liu00a}, violations of 4D causality in
the brane world \cite{Kal98,Kal00,Chu00,Ish01}, the tendency of
the brane world to attract or repel buck geodesics
\cite{Mar00,Mar01,Cha01}, the subtle interplay between 4D and 5D
affine parameters \cite{Cha00,Sea01a}, and the dynamical
acquisition and variation of particle rest masses from
dimensional reduction \cite{Mas94,Liu00b,Sea01a,Bil01}. In work
closely related to the study of higher dimensional geodesics,
several authors have considered the behaviour of pointlike
gyroscopes moving in 5D manifolds
\cite{Kal94,Mas98,Liu00a,Liu00b}.

We will not address all of these issues in this paper.  Rather,
we present a generic formalism which may aid in the study of
higher dimensional particle motion.  In particular, we wish to
describe the motion of test particles in a $(3+1)+1$ dimensional
manifold in a geometric and covariant manner.  Our approach is
inspired by the familiar geometric construction used in the
Arnowitt-Deser-Misner (ADM) Hamiltonian formulation of general
relativity, where the 4D manifold is foliated by a series of 3D
spacelike hypersurfaces $\Sigma_t$ \cite{Mis70,Wal84}.  As
described in detail in Section \ref{sec:geometry}, we introduce a
similar foliation of the 5D manifold by a series of timelike 4D
hypersurfaces $\Sigma_\ell$, each of which corresponds to a 4D
spacetime.  Here, $\ell$ labels the fifth dimension.  The beauty
of this approach is that it allows us to covariantly decompose 5D
tensors in terms of quantities either tangent or orthogonal to
$\Sigma_\ell$, and to derive equations that transform correctly
under 4D and 5D coordinate changes.

Before describing our geometric framework in detail in Section
\ref{sec:geometry}, we would like to outline the three main
motivations for this work.  First, we note that the majority of
the references cited above deal with higher dimensional geodesics
in 5D via the well known equation
\begin{equation}\label{standard geodesic}
    \frac{d^2 x^A}{d\lambda^2} + \Gamma^{A}_{BC} \frac{dx^B}{d\lambda}
    \frac{dx^C}{d\lambda} = 0,
\end{equation}
where $\lambda$ is a 5D affine parameter and $\Gamma^{A}_{BC}$
refers to the higher dimensional Christoffel
symbols.\footnote{See the note at the end of this section for an
accounting of the conventions used in this paper.}  Particle
trajectories are decomposed into a 4D part by considering the $A
= 0,1,2,3$ components of (\ref{standard geodesic}) differently
from the $A=4$ component, which governs the motion in the extra
dimension. The 4D part can then be manipulated to look like the
4D equation of motion of a particle subjected to a
non-gravitational acceleration; i.e.\ the ``fifth force''.  As
pointed out by Ponce de Leon, this procedure has some drawbacks
\cite{Pon01a,Pon01b}. The most serious one is that the algorithm
is not covariant in 5D, which results in the gauge dependence of
the extra force. Another problem is that the fifth force is not
orthogonal to the 4-velocity, which contradicts standard 4D
physics.  To overcome these difficulties, Ponce de Leon has
analyzed the 5D geodesic equation in terms of local basis vectors
and introduced suitable redefinitions of various quantities to
obtain a better-behaved description of test particle
trajectories.  We wish to present an alternative formalism based
on a $(3+1)+1$ decomposition of the 5D geodesic equation that is
more geometric than Ponce de Leon's algorithm, yet still corrects
the problems associated with descriptions based on (\ref{standard
geodesic}). This is the subject of Section \ref{sec:split}.

The second goal of this paper is to explore the confinement of
test particles to a given $\Sigma_\ell$ hypersurface.  An
essential ingredient of brane world models is that matter fields
are confined to a 3-brane that is identified with our universe.
In string theory, this confinement comes about naturally from
strings with endpoints attached to the brane.  In the RS picture,
localization of the zero-mode of the graviton and other fields
arises from the discontinuous nature of the geometry.  Other
authors have considered ``smoothed-out'' versions of the RS model
by introducing finite-width, or ``thick'', branes created via
dynamical degrees of freedom \cite{Csa00,Emp01,Gio01,Rog01}.
Localization of gravity and other fields comes from a steep, but
smoothly differentiable, extra-dimensional potential. In this
paper, we aim to explore the classical analogue of these, and
other, confinement mechanisms making use of the geometric
framework discussed above. Several workers have both noted that
in general, non-gravitational forces are required to keep test
particles confined to a thin 3-brane \cite{Bil97,Mar00,Ish01}. In
Section \ref{sec:confinement}, we will derive the general form of
such a confining force for a 3-brane of arbitrary geometry and
show that it is nothing but a generalization of the centripetal
force from undergraduate mechanics.

Our final motivation comes from a result derived in Section
\ref{sec:confinement}, which states that test particles confined
to a $\Sigma_\ell$ hypersurface by normal forces travel on
geodesics of that hypersurface.  This makes it impossible to
distinguish certain higher dimensional scenarios from ordinary
general relativity based on the kinematic data of test bodies.
This causes us to search for another test for non-compact extra
dimensions, namely the behaviour of pointlike gyroscopes in 5D
manifolds.  We extend previous work by describing the dynamics of
a spinning body in terms of the foliation language of Section
\ref{sec:geometry} and by considering both freely falling and
confined gyroscopes. In Section \ref{sec:spin}, we discuss the
general formulation of the point gyroscope problem in 5D.  In
Section \ref{sec:fermi}, we apply our splitting algorithm to the
spin equations of motion. In Section \ref{sec:brane}, we discuss
how our results should be interpreted in the presence of a thin
3-brane.  Finally, in Section \ref{sec:example}, we give an
example of how a point gyroscope behaves in a simple 5D
cosmological setting. Our main result is that when observed in
4D, the spin angular momentum of a gyroscope confined to a
$\Sigma_\ell$ hypersurface is not conserved and that the spin
tensor will precess with respect to a 4D non-rotating frame due
to the existence of an anomalous torque. Such effects may be
observable by space-based experiments such as Gravity Probe B
\cite[review]{Jan96}.

Our results are summarized in Section \ref{sec:conclusions} along
with suggestions for future work.

\paragraph*{Conventions}
The 5D metric signature is $(+----)$ while the 4D metric
signature is $(+---)$; i.e.\ the extra dimension is assumed to be
spacelike.  In this paper, uppercase Latin indices run from 0 --
4 and lowercase Greek indices run 0 -- 3. Early lowercase Latin
indices run over all four spatial dimensions 1 -- 4, while late
indices run over the three spatial dimensions 1 -- 3 associated
with the 4D manifold.

\section{Geometric Construction}\label{sec:geometry}

In this section, we introduce our foliation of the 5D manifold
and the various geometric quantities that figure prominently in
subsequent calculations.  Consider a 5D manifold $(M,g_{AB})$
with one timelike dimension and covered by an arbitrary system of
coordinates $x^A$.  We introduce a scalar function
\begin{equation}
    \ell = \ell(x^A),
\end{equation}
which is defined everywhere on $M$.  The 4D hypersurfaces $\ell$
= constant, which we shall denote by $\Sigma_\ell$, are assumed
to have a spacelike normal vector field given by
\begin{equation}
    n_A = \Phi \di_A \ell, \quad n^A n_A = -1.
\end{equation}
Each of the $\Sigma_\ell$ hypersurfaces may be associated with a
3+1 dimensional spacetime. The scalar $\Phi$ which normalizes
$n^A$ is known as the lapse function.  We define the projection
tensor as
\begin{equation}
\label{induced metric}
    h_{AB} = g_{AB} + n_A n_B.
\end{equation}
This tensor is symmetric $h_{AB} = h_{BA}$ and orthogonal to
$n_A$.  We place a 4D coordinate system on each of the
$\Sigma_\ell$ hypersurfaces $y^\alpha$.  The four basis vectors
\begin{equation}
    e^A_\alpha = \frac{\di x^A}{\di y^\alpha}, \quad n_A
    e^A_\alpha = 0
\end{equation}
are by definition tangent the $\Sigma_\ell$ hypersurfaces and
orthogonal to $n^A$.  It is easy to see that $e^A_\alpha$ behaves
as a 5D contravariant vector under 5D coordinate transformations
$x^A \rightarrow \bar{x}^A$ and a 4D covariant vector under 4D
coordinate transformations $y^\alpha \rightarrow \bar{y}^\alpha$.
We can use these basis vectors to project 5D objects onto
$\Sigma_\ell$ hypersurfaces.  For example, for an arbitrary 5D
covariant vector
\begin{equation}
    T_\alpha = e_\alpha^A T_A.
\end{equation}
Here, $T_\alpha$ is said to be the projection of $T_A$ onto
$\Sigma_\ell$.  Clearly, $T_\alpha$ behaves as a scalar under 5D
coordinate transformations and a covariant vector under 4D
transformations.  The induced metric on the $\Sigma_\ell$
hypersurfaces is given by
\begin{equation}
    h_{\alpha\beta} = e^A_\alpha e^B_\beta g_{AB} = e^A_\alpha
    e^B_\beta h_{AB}.
\end{equation}
Just like $g_{AB}$, the induced metric has an inverse
\begin{equation}
    h^{\alpha\gamma} h_{\gamma\beta} = \delta^{\alpha}_{\beta}.
\end{equation}
The induced metric and its inverse can be used to raise and lower
indices of 4-tensors and change the position of the spacetime
index of the $e^A_\alpha$ basis vectors.  This then implies
\begin{equation}
    e_A^\alpha e^A_\beta = \delta^\alpha_\beta.
\end{equation}
Also note that since $h_{AB}$ is entirely orthogonal to $n^A$, we
can express it as
\begin{equation}\label{induced decomposition}
    h_{AB} = h_{\alpha\beta} e^\alpha_A e^\beta_B.
\end{equation}

We will also require an expression that relates 5D covariant
derivative of 5-tensors to the 4D covariant derivative of the
corresponding 4-tensors.  For covariant 5-vectors, the
appropriate formula is
\begin{equation}
    \nabla_\beta T_\alpha = e^B_\beta e^A_\alpha \nabla_B
    (h_A{}^C T_C),
\end{equation}
where $\nabla_B$ is the 5D covariant derivative on $M$ defined
with respect to $g_{AB}$ and $\nabla_\beta$ is the 4D covariant
derivative defined with respect to $h_{\alpha\beta}$.  The
generalization to tensors of higher rank is obvious.  It is not
difficult to confirm that this definition of $\nabla_\alpha$
satisfies all the usual requirements imposed on the covariant
derivative operator.

At this juncture, it is convenient to introduce our definition of
the extrinsic curvature $K_{\alpha\beta}$ of the $\Sigma_\ell$
hypersurfaces:
\begin{equation}\label{extrinsic def}
    K_{\alpha\beta} = e^A_\alpha e^B_\beta \nabla_A n_B =
    \tfrac{1}{2} e^A_\alpha e^B_\beta \pounds_n h_{AB}.
\end{equation}
Note that the extrinsic curvature is symmetric (i.e.\
$K_{\alpha\beta} = K_{\beta\alpha}$) and may be thought of as the
derivative of the induced metric in the normal direction.  This
4-tensor will appear often in what follows.

Finally, we note that $(y^\alpha,\ell)$ defines an alternative
coordinate system to $x^A$.  The appropriate diffeomorphism is
\begin{equation}\label{diffeo}
    dx^A = e_\alpha^A dy^\alpha + \ell^A d\ell,
\end{equation}
where
\begin{equation}\label{flow of l vector}
    \ell^A =  \left( \frac{\di x^A}{\di \ell} \right)_{y^\alpha =
    \mathrm{const.}}
\end{equation}
is the vector tangent to lines of constant $y^\alpha$.  We can
always decompose 5D vectors into the sum of a part tangent to
$\Sigma_\ell$ and a part normal to $\Sigma_\ell$.  For $\ell^A$
we write
\begin{equation}\label{l vector}
    \ell^A = N^\alpha e_\alpha^A - \Phi n^A.
\end{equation}
This is consistent with $\ell^A \di_A \ell = 1$, which is
required by the definition of $\ell^A$, and the definition of
$n^A$.  The 4-vector $N^\alpha$ is called the shift vector and it
describes how the $y^\alpha$ coordinate system changes as one
moves from a given $\Sigma_\ell$ hypersurface to another.  Using
our formulae for $dx^A$ and $\ell^A$, we can write the 5D line
element as
\begin{eqnarray}\nonumber
    d\mathcal{S}^2 & = & g_{AB} \, dx^A dx^B \\ & = & h_{\alpha\beta}
    (dy^\alpha + N^\alpha d\ell) (dy^\beta + N^\beta d\ell)
    - \Phi^2 \, d\ell^2,
\end{eqnarray}
which reduces to $d\mathcal{S}^2 = h_{\alpha\beta} dy^\alpha
dy^\beta$ if $d\ell = 0$.

This completes our description of the geometric structure we will
use in the rest of the paper.  We would like to stress that this
formalism does not depend on the form of the higher dimensional
field equations, or the choice of 5D or 4D coordinates.  It is
sufficiently general to be applied to a wide class of $(3+1)+1$
theories of gravity.

\section{Splitting of the 5D Geodesic Equation}\label{sec:split}

\subsection{5D Affine Parameterization}\label{sec:affine}

In this section, we utilize the formalism introduced above to
split the 5D geodesic equation into a series of relations
involving quantities that are either tangent or orthogonal to the
$\Sigma_\ell$ hypersurfaces.  The goal is to derive the 4D
equation of motion, an equation governing the motion in the extra
dimension, and an equation governing the norm of the 4-velocity.

We consider a 5D geodesic trajectory with 5-velocity $u^A$
satisfying
\begin{subequations}\label{original equations}
\begin{eqnarray}\label{5D geodesic} 0 & = & u^A
    \nabla_A u^B, \\ \label{5D norm} \epsilon & = & u^A u_A,
\end{eqnarray}
\end{subequations}
with
\begin{equation}\label{u def}
    u^A = \dot{x}^A,
\end{equation}
where an overdot denotes $D/d\lambda = u^A \nabla_A$ and
$\epsilon = 0,\pm 1$. Since the norm of $u^A$ is constant,
$\lambda$ is an affine parameter. We define
\begin{equation}
    \xi \equiv n_A u^A,
\end{equation}
which allows us to write
\begin{subequations}\label{u decomposition}
\begin{eqnarray}\label{first u decomposition}
    u^A & = & h^{AB} u_B - \xi n^A \\ \label{second u decomposition}
    & = & e^A_\alpha u^\alpha - \xi n^A.
\end{eqnarray}
\end{subequations}
using (\ref{induced metric}) and defining $u^\alpha = e^\alpha_A
u^A$.  Putting (\ref{first u decomposition}) into (\ref{5D
geodesic}) and expanding yields
\begin{equation}\label{first decomposition}
    0 = h^{AC} u_C \nabla_A ( h^{BM} u_M - \xi n^B) - \xi n^A
    \nabla_A u^B.
\end{equation}
Contracting this with $e_B^\beta$ and using the fact that $h^{AC}
= e^A_\alpha e^C_\gamma h^{\alpha\gamma}$ and $e_B^\beta n^B =
0$, we get
\begin{equation}
    u^\alpha \nabla_\alpha u^\beta = \xi (K^{\alpha\beta} u_\alpha +
    e^\beta_B n^A \nabla_A u^B),
\end{equation}
where $K_{\alpha\beta}$ is defined by equation (\ref{extrinsic
def}).

Returning to (\ref{5D geodesic}) and (\ref{first u
decomposition}) we can write
\begin{equation}\label{second decomposition}
    0 = (h^{AM} u_M - \xi n^A) \nabla_A (h^{BC} u_C) - u^A
    \nabla_A (\xi n^B)
\end{equation}
instead of (\ref{first decomposition}).  We can contract this
with $n_B$ and use the facts that
\begin{eqnarray}\nonumber
    0 = h_{BC} n^B u^C & \Rightarrow & n_B \nabla_A (h^{BC} u_C) =
    - h^{BC} u_C \nabla_A n_B \\ \nonumber
    -1 = n_A n^A & \Rightarrow & n_B \nabla_A n^B = 0,
\end{eqnarray}
to obtain, after some algebra
\begin{equation}
    \dot{\xi} = K_{\alpha\beta} u^\alpha u^\beta - \xi n^A
    u^B \nabla_A n_B,
\end{equation}
where we have noted that $\dot{\xi} = d\xi/d\lambda = u^A
\nabla_A \xi$.

Continuing, we note that $\epsilon = g_{AB} u^A u^B$ can be
expanded by making the substitution $g_{AB} = h_{\alpha\beta}
e^\alpha_A e^\beta_B - n^A n^B$.  The result is
\begin{equation}
    \epsilon = h_{\alpha\beta} u^\alpha u^\beta - \xi^2.
\end{equation}

In summary, equations (\ref{original equations}) can be rewritten
as
\begin{subequations}\label{split equations}
\begin{eqnarray}\label{4D part}
    a^\beta(u) & = & \xi (K^{\alpha\beta} u_\alpha +
    e^\beta_B n^A \nabla_A u^B), \\ \label{extra part}
    \dot{\xi} & = & K_{\alpha\beta} u^\alpha u^\beta - \xi n^A
    u^B \nabla_A n_B, \\ \label{split norm}
    \epsilon & = & h_{\alpha\beta} u^\alpha u^\beta - \xi^2.
\end{eqnarray}
\end{subequations}
Here, we have defined the acceleration of a 4-vector by
\begin{equation}\label{define acceleration}
    a^\beta(q) \equiv q^\alpha \nabla_\alpha q^\beta.
\end{equation}
It must be noted that these equations do not represent a strict
$(3+1)+1$ splitting of the geodesic equation because the 5D
vector $u^A$ appears on the righthand side of equations (\ref{4D
part}) and (\ref{extra part}).  This shortcoming can be easily
alleviated by making use of (\ref{second u decomposition}), but
we find that the present form of the equations is more useful for
subsequent calculations. We will therefore abstain from further
manipulations.

As a consistency check, we can contract (\ref{4D part}) with
$u_\beta$.  After some algebra, we obtain
\begin{equation}
    u^\alpha \nabla_\alpha (u^\beta u_\beta) = 2\xi (
    K_{\alpha\beta} u^\alpha u^\beta + h_{BC} u^C n^A \nabla_A
    u^B).
\end{equation}
Substituting $h_{BC} = g_{BC} + n_B n_C$ and using equation
(\ref{split norm}) yields after further manipulation
\begin{equation}
    u^\alpha \nabla_\alpha \xi - \xi n^A \nabla_A \xi =
    K_{\alpha\beta} u^\alpha u^\beta - \xi n^A u^B \nabla_A n_B.
\end{equation}
But, the lefthand side is easily seen to be equivalent to $u^A
\nabla_A \xi = \dot{\xi}$.  Hence, it is possible to derive
equation (\ref{extra part}) from equations (\ref{4D part}) and
(\ref{split norm}).  Therefore, equations (\ref{split equations})
are mutually consistent in that one of the set is redundant.

In conclusion, we have succeeded in splitting the 5D geodesic
equation (\ref{5D geodesic}) into an equation of motion for the
4-velocity (\ref{4D part}) and an equation governing motion in
the extra dimension (\ref{extra part}).  We have also rewritten
the 5D affine parameter condition (\ref{5D norm}) in a matter
consistent with the $(3+1)+1$ split (\ref{split norm}).

\subsection{Parameter Transformation to 4D Proper
Time}\label{sec:proper}

Upon examination of equations (\ref{split equations}), it becomes
clear that the 5D affine parameter $\lambda$ cannot be the same
as what is usually called the 4D proper time $s$.  The reason is
that the norm of the 4-velocity $u^\alpha$ is not equal to unity
by equation (\ref{split norm}).  In this section, we will detail
a parameter transformation from the 5D affine parameter $\lambda$
to the 4D proper time $s$ that will make the norm of the
4-velocity constant.

Our parameter transformation is described by the following
formulae:
\begin{equation}
    u^A = \psi v^A, \quad \psi = \frac{ds}{d\lambda}, \quad v^A =
    \frac{dx^A}{ds}.
\end{equation}
$v^A$ is the 5-velocity of the test particle in the
$s$-parameterization.  We need to also define
\begin{equation}
    \chi = n^A v_A \Rightarrow \xi = \chi\psi.
\end{equation}
The equation on the right follows from $\xi = n^A u_A$.  We can
substitute these expressions into equations (\ref{split
equations}) in order to see what the $(3+1)+1$ split of the
geodesic equation looks like in the $s$ parameterization.  After
a straightforward, but tedious, calculation, we obtain
\begin{subequations}\label{transformed split equations}
\begin{eqnarray}\nonumber
    v^\alpha \nabla_\alpha v^\beta & = & \chi (K^{\alpha\beta}
    v_\alpha +
    e^\beta_B n^A \nabla_A v^B) \\ \label{transformed 4D part} & &
    - v^\beta (\ln\psi)', \\
    \label{transformed extra part}
    d(\chi\psi)/ds & = & \psi( K_{\alpha\beta} v^\alpha v^\beta
    - \chi n^A
    v^B \nabla_A n_B), \\ \label{transformed split norm}
    \epsilon & = & \psi^2 (h_{\alpha\beta} v^\alpha v^\beta
    - \chi^2).
\end{eqnarray}
\end{subequations}
Here, a prime $'$ denotes $D/ds = v^A \nabla_A$.  In order to go
further, we need to demand that the norm of the 4-velocity be
unity in the proper time parameterization:
\begin{subequations}\label{s conditions}
\begin{eqnarray}
    1 & = & h_{\alpha\beta} v^\alpha v^\beta, \\ \label{s condition 2}
    \epsilon & = & \psi^2 (1 - \chi^2),
\end{eqnarray}
\end{subequations}
where the bottom equation follows from (\ref{transformed split
norm}).  We can use (\ref{s condition 2}) to show $(\chi\psi)' =
\psi'/\chi$, which may then be substituted into equation
(\ref{transformed extra part}) to isolate $(\ln\psi)'$. The
resulting formula can then be inserted into (\ref{transformed 4D
part}).  This penultimate expression may be simplified by the use
of the identity
\begin{equation}
    \chi n^A v^B \nabla_A n_B = -h_{BC} n^A v^C \nabla_A v^B,
\end{equation}
which is obtained by operating $n^A \nabla_A$ on both sides of $1
= h_{BC} v^B v^C$ and then using equation (\ref{induced metric}).
We finally get the following expression for the 4-acceleration of
$v^\alpha$:
\begin{eqnarray}\nonumber
    a^\beta(v) & = & \chi \Big[ ( K^{\alpha\beta} v_\alpha -
    K_{\mu\nu} v^\mu v^\nu v^\beta ) + \\ \label{s acceleration}
    & & ( e_B^\beta - e_B^\gamma v_\gamma v^\beta) n^A \nabla_A
    v^B \Big].
\end{eqnarray}
Note that this acceleration is orthogonal to $v^\alpha$,
\begin{equation}
    v_\beta a^\beta(v) = 0,
\end{equation}
which also follows from $v^\alpha v_\alpha = 1$.  This is in
contrast to 5D affine parameterization, which has
\begin{equation}
    u_\beta a^\beta(u) \ne 0.
\end{equation}
That is, in the proper time parameterization the 4-acceleration
is orthogonal to the 4-velocity, while in the 5D affine
parameterization the 4-acceleration has components parallel to
the 4-velocity.

To complete our discussion of the proper time parameterization,
we need to specify how the velocity along the extra dimension
$\chi$ evolves with $s$.  Going back to equation (\ref{s
condition 2}), it is clear that if $\epsilon = 0$ then $\chi =
\pm 1$.  That is, if the 5D path is null, we must have $n^A v_A =
\pm 1$.  It is interesting to note that even if the 5D path is a
null geodesic with $v^A v_A = 0$, we can still have the 4D
trajectory as timelike with $v^\alpha v_\alpha = 1$.  This
correspondence between massless trajectories in 5D and massive
ones in 4D has been noted before \cite{Sea01a,Bil01,You01}.  For
cases where $\epsilon = \pm 1$, we can use (\ref{s condition 2})
to get $(\chi\psi)' = \psi \chi' / (1-\chi^2)$.  This into
(\ref{transformed extra part}) gives
\begin{eqnarray}\nonumber
\chi' & = & (1 - \chi^2) ( K_{\alpha\beta} v^\alpha v^\beta -
\chi n^A v^B \nabla_A n_B), \quad \epsilon = \pm 1, \\ \chi & = &
\pm 1, \quad \epsilon = 0. \label{s extra}
\end{eqnarray}
These formulae, along with (\ref{s acceleration}) and equations
(\ref{s conditions}) give our description of the $(3+1)+1$ split
of the 5D geodesic equation in the proper time parameterization.

\subsection{Differences between the current formalism and the
literature} \label{sec:differences}

Before discussing the confinement of particle trajectories to a
given hypersurface, we shall discuss some of what makes the
current work different from previous studies.  As mentioned in
Section \ref{sec:introduction}, a number of authors have
considered the force-free motion of particles in 5D, but most
have concentrated on the determination of equations of motion
from (\ref{standard geodesic}) or the equivalent variational
principle \cite[references therein]{Wes01a}. The splitting of
spacetime from the extra coordinate has been achieved by
considering the first four equations of motion differently from
the last one. The algorithm presented in Section \ref{sec:split}
achieves the splitting in a more geometric fashion, employing the
$(3+1)+1$ dimensional foliation technology introduced in Section
\ref{sec:geometry}. This fundamental difference in methodology
results in two main differences between descriptions of 5D
geodesics.  We proceed to outline these differences in this
section.

First, we note that the central object in our description of
particle trajectories is $u^\alpha = e^\alpha_A u^A$ while in the
literature it is $\dot{y}^\alpha =
dy^\alpha/d\lambda$.\footnote{For concreteness, we limit
ourselves to the $\lambda$ parameterization, although our
comments apply equally well to the $s$ parameterization.}  But
these are not equal.  To see this, we can use equations
(\ref{diffeo}), (\ref{l vector}) and (\ref{u def}) to get
\begin{equation}
    u^A = e^A_\alpha (\dot{y}^\alpha + \dot{\ell} N^\alpha) - \Phi
    \dot{\ell} n^A.
\end{equation}
This then yields
\begin{subequations}\label{4D u and xi}
\begin{eqnarray}\label{4D u}
    u^\alpha & = & \dot{y}^\alpha + \dot{\ell} N^\alpha, \\
    \label{4D xi} \xi & = & \Phi \dot{\ell}.
\end{eqnarray}
\end{subequations}
These equations essentially replace equation (\ref{u def}) in the
same manner that equations (\ref{split equations}) replace
equations (\ref{original equations}).  The important thing to
note is that $u^\alpha$ is not equal to $dy^\alpha / d\lambda$,
as may na\"{\i}vely be assumed.  We can understand equation
(\ref{4D u}) by interpreting it as an equation concerning
relative velocities.  Geometrically, equation (\ref{flow of l
vector}) tells us that the $\dot{\ell} N^\alpha$ term in (\ref{4D
u}) represents the projected velocity of points on the $y^\alpha$
coordinate grid with respect to a ``stationary'' coordinate frame
$\bar{y}^\alpha$ (i.e.\ a system of coordinates on $\Sigma_\ell$
with $N^\alpha = 0$).  This stationary coordinate frame has been
extensively studied in the literature and has been termed
``canonical'' \cite{Mas98,Sea01a,Bil01}.\footnote{The warped
product metric ansatz popular in brane world models is an example
of a canonical coordinate system.} Obviously, $\dot{y}^\alpha$ is
the 4-velocity of the particle with respect to the $y^\alpha$
grid. Therefore, $u^\alpha$ is the velocity of the particle with
respect to the $\bar{y}^\alpha$ frame, or, in other words, the
velocity of the particle in canonical coordinates.  The fact that
$u^\alpha \ne \dot{y}^\alpha$ is not particularly worrisome
because having solved equations (\ref{split equations}) it is
easy to obtain $\dot{y}^\alpha$ and $\dot{\ell}$ from equations
(\ref{4D u and xi}).

The second way in which the results of Sections \ref{sec:affine}
and \ref{sec:proper} differ from previous studies has to do with
the presentation of the equations of motion.  In the literature,
higher dimensional geodesics are often analyzed in terms of the
so-called ``fifth force'', which for any 4-vector $q^\alpha$ is
defined as
\begin{equation}\label{f definition}
    f^\alpha(q) = \frac{dq^\alpha}{d\lambda} +
    \Gamma^\alpha_{\beta\gamma} q^\beta q^\gamma,
\end{equation}
where $\Gamma^\alpha_{\beta\gamma}$ are the Christoffel symbols
associated with $h_{\alpha\beta}$.  In 4D, $f^\alpha(q)$ is
identical to the acceleration of $q^\alpha$.  However, the
equality between $f^\alpha(q)$ and $a^\alpha(q)$ does not hold in
5D. To see this, we write
\begin{eqnarray}
    \nonumber f^\alpha(q) & = & q^A \di_A q^\alpha +
    \Gamma^\alpha_{\beta\gamma} q^\beta q^\gamma \\ \nonumber
    & = & (h^{AB} - n^A n^B) q_A \di_B q^\alpha +
    \Gamma^\alpha_{\beta\gamma} q^\beta q^\gamma \\ \nonumber & =
    & q^\beta\di_\beta q^\alpha +  \Gamma^\alpha_{\beta\gamma}
    q^\beta q^\gamma - (q_A n^A) n^B \di_B q^\alpha \\ & = & q^\beta
    \nabla_\beta q^\alpha - (q_A n^A) n^B \di_B q^\alpha.
\end{eqnarray}
In going from the second to the third line, we have used the fact
$e^B_\beta \di_B = \di_\beta$ after making the substitution
$h^{AB} = h^{\alpha\beta} e^A_\alpha e^B_\beta$.  We have
therefore established that fifth force is not equal to the 4D
acceleration vector, instead they are related via
\begin{equation}\label{f to a transformation}
    f^\alpha(q) = a^\alpha(q) - (q_B n^B) n^A \di_A q^\alpha.
\end{equation}
This equation raises an important point about the behaviour of
$f^\alpha$ under 4D coordinate transformations.  It is obvious
from equation (\ref{define acceleration}) that $a^\alpha(q)$ is a
4-vector.  But we will now demonstrate that $- \xi n^A \di_A
q^\alpha$ is not.  Consider a 4D coordinate transformation
$y^\alpha \rightarrow \tilde{y}^\alpha$. Under such a
transformation, we know that $q^\alpha$ transforms as a 4-vector:
$\tilde{q}^\alpha = (\di \tilde{y}^\alpha / \di y^\beta)
q^\beta$. This implies the following transformation law for $n^A
\di_A q^\alpha$:
\begin{equation}
    n^A \di_A \tilde{q}^\alpha = \frac{\di\tilde{y}^\alpha}
    {\di y^\beta} n^A \di_A q^\beta + \frac{q^\beta}{\Phi}
    \left( N^\mu \di_\mu - \di_\ell \right)
    \frac{\di\tilde{y}^\alpha} {\di y^\beta}.
\end{equation}
Here, we have used equation (\ref{l vector}) to substitute for
$n^A$ and then the definitions of $e^A_\alpha$ and $\ell^A$ with
the chain rule to transform the partial derivatives.  The first
term on the RHS is what one would expect to see if $n^A \di_A
q^\alpha$ was indeed a 4-vector.  But the presence of the second
term indicates that it is not.  In particular, if either the
shift vector $N^\alpha$ is nonzero or the 4D coordinate
transformation depends on $\ell$, then $n^A \di_A q^\alpha$ will
not satisfy the usual tensor transformation law.  This of course
means that the fifth force defined by (\ref{f definition}) is not
a 4-vector.  Also, we note that several authors have found that
in the $s$-parameterization the fifth force is not orthogonal to
4-velocity, which is in direct contradiction with standard 4D
physics.  This difficulty is removed by adopting the description
of Section \ref{sec:proper}, where the acceleration of $v^\alpha$
is properly orthogonal to $v^\alpha$.  For these two reasons, we
prefer to describe the 5D geodesics in terms of $a^\alpha$ as
opposed to $f^\alpha$.  This choice is not critical because one
can easily move between the two descriptions by using equation
(\ref{f to a transformation}).

\section{Confinement of trajectories to $\Sigma_\ell$
hypersurfaces}\label{sec:confinement}

As mentioned in Section \ref{sec:introduction}, a variety of
different mechanisms have been suggested to confine matter fields
to a 3-brane.  In this section, we explore the classical particle
analogue of this field concept.  Our goal is to find out what
kind of force per unit mass is required to confine test particles
to a given $\Sigma_\ell$ hypersurface, and to determine the form
of the 4D equation of motion.  Our description is essentially
that of an effective theory, since we do not speculate about the
source of this non-gravitational acceleration.  In what follows,
we will use the term force to refer to what should properly be
called a force per unit mass.

There are several possible avenues one can use to derive the
confinement force.  One possibility is to modify the calculation
of Section \ref{sec:affine} to include an undetermined external
force.  Then, one can enforce the confinement of the particle by
demanding $\xi = 0$, which in turn places constraints on the
external force. Another route begins with the Gauss-Weingarten
equation, which relates the 4D and 5D accelerations of a curve
confined to a given $\Sigma_\ell$ hypersurface, and can hence be
used to fix the form of the non-gravitational force
\cite{Mis70,Ish01}. However, if one wishes to proceed from first
principles, a particularly transparent derivation comes from the
method of Lagrange multipliers, which is what we will give in
this section.

We take the 5D particle Lagrangian, in the affine
parameterization, to be
\begin{equation}
    \mathcal{L} = \tfrac{1}{2} g_{AB} u^A u^B + \mu( \lambda )
    \Phi (  \lambda ) \left[ \ell(x^N) - \ell_0 \right].
\end{equation}
Here, the constraint on the particle motion is given by
$\varphi(x^A) = \ell(x^A) - \ell_0 = 0$, which essentially means
that the trajectory is confined to the $\Sigma_\ell$ hypersurface
corresponding to $\ell(x^A) = \ell_0$.  The undetermined function
$\mu(\lambda)$ is the Lagrange multiplier.  We have factored out
a $\Phi(\lambda)$ term, which is the lapse function evaluated
along the trajectory.  We have done this to simplify the
equations of motion, which are obtained from the standard
Euler-Lagrange formulae. This result is
\begin{equation}\label{Lagrange EOM}
    u^B \nabla_B u^A = \mu n^A,
\end{equation}
where we have used $n_A = \Phi \di_A \ell$.  Now, because
$\varphi = 0$ along the trajectory, we require that $u^A \di_A
\varphi = 0$.  This condition may be written as
\begin{equation}\label{constraint}
    0 = \xi = u^A n_A,
\end{equation}
which is an obvious requirement for paths confined to a given
$\Sigma_\ell$ hypersurface.  We now contract both sides of
(\ref{Lagrange EOM}) with $n_A$ and make use of the fact that
(\ref{constraint}) implies that $n_A \nabla_B u^A = - u^A
\nabla_B n_A$ to obtain
\begin{equation}
    \mu = u^A u^B \nabla_A n_B.
\end{equation}
Finally, we note that since $\xi = 0$ we can write $u^A =
e^A_\alpha u^\alpha$, which yields that
\begin{equation}\label{mu solution}
    \mu = K_{\alpha\beta} u^\alpha u^\beta.
\end{equation}
This result fixes the force of constraint $\mu n^A$ that appears
on the RHS of the equation of motion (\ref{Lagrange EOM}).

We now wish to address the issue of what happens to the $(3+1)+1$
splitting performed in Section \ref{sec:affine} in the presence
of this confinement force.  We can replace the ordinary geodesic
equation (\ref{5D geodesic}) with the constrained equation of
motion (\ref{Lagrange EOM}) and our solution for $\mu$.  Then,
without first demanding that $\xi = 0$, we can redo the
manipulations of Section \ref{sec:affine} on the new equations.
It transpires that equations (\ref{4D part}) and (\ref{split
norm}) are unaffected by the presence of the confining force.
However, equation (\ref{extra part}) is modified to read
\begin{equation}\label{modified extra part}
    \dot{\xi} = - \xi n^A u^B \nabla_A n_B.
\end{equation}
One possible solution to the modified system of equations formed
by (\ref{4D part}), (\ref{split norm}), and (\ref{modified extra
part}) is $\xi = 0$.  Assuming that we do have a situation where
$\xi = 0$, then spilt geodesic equations of motion given by
(\ref{split equations}) become
\begin{equation}\label{confined equations}
    u^\beta \nabla_\beta u^\alpha = 0, \quad \xi = 0, \quad \epsilon =
    h_{\alpha\beta} u^\alpha u^\beta.
\end{equation}
In other words, we have discovered that if the higher dimensional
equation of motion is given by
\begin{equation}\label{5D confined}
    u^B \nabla_B u^A = ( K_{\alpha\beta} u^\alpha u^\beta ) n^A,
\end{equation}
and we impose the initial condition $\xi = 0$, then the particle
trajectory will be confined to a single $\Sigma_\ell$
hypersurface.  In addition, the particle will travel on a 4D
geodesic of that hypersurface, defined by $u^\beta \nabla_\beta
u^\alpha = 0$ and $u^\alpha u_\alpha = \epsilon$.  In more
physical terms, we can say that motion of the particle under the
action of the 5D confinement force looks like force-free 4D
motion on $\Sigma_\ell$.

This conclusion merits a few comments before we move on to the
next section.  First, the form of the confinement force could
have been anticipated from elementary physics.  Although our
formulae have been derived with a 5D manifold in mind, they hold
equally well in any dimension.  So, consider a 2+1 dimensional
flat manifold in polar coordinates with a line element
\begin{equation}
    ds^2 = dt^2 - dr^2 - r^2 d\phi^2.
\end{equation}
Suppose that in this manifold there is a particle confined to
move on an $r = R$ hypersurface; i.e.\ on a circle of radius $R$.
Then, the force per unit mass constraining the trajectory has a
magnitude of $|K_{\alpha\beta} u^\alpha u^\beta| = R (d\phi/ds)^2
= v^2/R$, where $v = R \, d\phi/ds$ is the linear spatial
velocity.  This result is recognized as the centripetal
acceleration of a particle moving in a circle from undergraduate
mechanics. Therefore, the confinement force we have derived in
this section is nothing more than the higher dimensional
generalization of the familiar centripetal acceleration.

Second, the causal properties of the 5D trajectories are
preserved when they are confined to 4D hypersurfaces.  This is,
the fact that $u^A u_A = u^\alpha u_\alpha = \epsilon$ implies
that timelike paths in 5D remain timelike when confined to
$\Sigma_\ell$, null paths in 5D remain null when confined to
$\Sigma_\ell$, etc\ldots.  This is contrast with Section
\ref{sec:proper}, where we saw that the projection of a 5D null
geodesic path onto a $\Sigma_\ell$ hypersurface could be
timelike, but with a complicated equation of motion (\ref{s
acceleration}).  In other words, free massless particles in 5D
can look like accelerated massive particles in 4D, but confined
massless particles in 5D look like free massless particles in 4D.
On a related note, the 5D affine parameter $\lambda$ coincides
with the 4D proper time $s$ for confined paths.

The third point is that the confining force vanishes if
$K_{\alpha\beta} = 0$.  In this case, geodesics on $\Sigma_\ell$
are automatically geodesics of the 5D manifold $M$.  As pointed
out by Ishihara, this is hardly a new result \cite{Ish01}.
Hypersurfaces that have $K_{\alpha\beta} = 0$ are known as
geodesically complete.  However, it should be pointed out that
$K_{\alpha\beta} = 0$ is a sufficient, but not necessary
condition for a geodesic on $\Sigma_\ell$ to also be a geodesic
of $M$.  The necessary condition is $K_{\alpha\beta} u^\alpha
u^\beta = 0$, which can be satisfied if $K_{\alpha\beta} \ne 0$.
For example, the asymptotes of a hyperboloid $S$ in Euclidean
3-space are geodesics of both $S$ and $\mathbb{E}_3$.  But, it is
not difficult to show that if all the geodesics on $\Sigma_\ell$
are geodesics of $M$, then $K_{\alpha\beta}$ is necessarily zero.

Fourth, we should comment on how the calculations of this section
fit into the brane world scenario.  The observant reader will
have already noticed that our manipulations implicitly assume
that the 5D manifold is smooth and free of defects.  Since this
is not the case in the thin brane world scenario, one may
legitimately wonder whether or not our results apply to a
particle confined to a 3-brane.\footnote{Of course, this is not
an issue in the thick brane world scenario, where the manifold is
smooth in the neighborhood of the the brane.}  To answer the
question, suppose that our 5D manifold contains a 3-brane
corresponding to the $\Sigma_{\ell = 0} = \Sigma_0$ hypersurface.
We can view the trajectory of a confined particle on the brane
$\gamma_0$ as the limit of a series of confined trajectories on
hypersurfaces living in the bulk.  To have a sensible theory, the
limiting procedure must result in the same $\gamma_0$ curve as
the brane is approached from the $\ell > 0$ and $\ell < 0$ sides.
Now, consider the curves $\gamma_{\pm}$ located on the
hypersurfaces $\Sigma_{\pm}$ at $\ell = 0^\pm$. Since the
$\Sigma_\pm$ hypersurfaces are in the bulk, $\gamma_\pm$ satisfy
equations (\ref{confined equations}) with $h_{\alpha\beta}
\rightarrow h_{\alpha\beta}^\pm$.  But one of Israel's junction
conditions, which must be satisfied in the neighborhood of the
brane, is
\begin{equation}
    [h_{\alpha\beta}] = 0,
\end{equation}
where we have used the usual jump notation: $[(\cdots)] =
(\cdots)_{\ell = 0^+} - (\cdots)_{\ell = 0^-}$.  That is, the
intrinsic 4-geometry of the $\Sigma_\ell$ hypersurfaces must be
continuous across the brane.  Since the confined trajectories
$u^{A}_\pm$ are determined entirely by the intrinsic geometry, we
see that both $\gamma_+$ and $\gamma_-$ must approach $\gamma_0$
as $\ell \rightarrow 0$.  Therefore, the confined trajectories on
the brane are perfectly well defined and are described by
(\ref{confined equations}).  However, it is interesting to note
that the acceleration of the $\gamma_\pm$ curves, as given in
equation (\ref{5D confined}), will not be continuous across the
brane.  This is because of the $\mathbb{Z}_2$ symmetry of the
brane world, which says the the extrinsic curvature of the
$\Sigma_\ell$ hypersurfaces is discontinuous about $\ell = 0$:
\begin{equation}\label{K jump}
    K_{\alpha\beta}^+ = -K_{\alpha\beta}^- \Rightarrow
    [K_{\alpha\beta}] \ne 0.
\end{equation}
Therefore, the 5D acceleration of $\gamma_0$ cannot be sensibly
defined because the one-sided limits of $A^B = K_{\alpha\beta}
u^\alpha u^\beta n^B$ do not agree.  At this juncture, this is
not a source of concern because $A^B$ is orthogonal to the brane
and is hence not directly measurable by observers.  However, we
shall see below that this acceleration is measurable in a
different context, which will necessitate careful consideration.

Our fifth, and final, point is that the 4D equation of motion
$u^\alpha \nabla_\alpha u^\beta = 0$ means that we cannot
observationally distinguish between a purely 4D universe with
potentially exotic matter and a brane world type scenario from
the kinematic data of macroscopic test particles.\footnote{We
exclude from the discussion possible short-range modifications of
Newton's gravitational law due to the 5D graviton propagator
because it is a quantum effect.}  In both cases, we have geodesic
motion on the 4D manifold. If we want to determine if our world
is fundamentally 4D or if we are merely confined to a 4D
hypersurface, we need to introduce new concepts, which is the
subject of the next section.

\section{5D pointlike gyroscopes}\label{sec:gyro}

\subsection{A spinning particle in 5D}\label{sec:spin}

In Section \ref{sec:confinement}, we saw that if a particle is
confined to a $\Sigma_\ell$ hypersurface by a centripetal
confinement force, then it will travel on a geodesic of
$\Sigma_\ell$.  This means that we cannot observationally
distinguish between confined motion in 5D and free motion in 4D
by studying the form of the trajectory $x^\alpha =
x^\alpha(\lambda)$.  However, just as an observer in a closed
vessel can use a gyroscope to determine if he is in a rotating
reference frame, we will see that we can use spinning bodies to
determine if apparently free particles are in actuality
accelerating in higher dimensions.

Our staring point is the equations of motion for a
``point-dipole'' spinning particle moving in a 5D manifold. These
equations for force free motion in 4D given by Papapetrou
\cite{Pap51,Mas71} and were later generalized by Schiff to
include non-gravitational forces and pointlike gyroscopes
\cite{Sch60a,Sch60b}. The extension to 5D is trivial, provided
that we assume that any non-gravitational forces exert no torque
on the body.  The equation of motion for the anti-symmetric spin
tensor $\sigma^{AB}$ is
\begin{equation}\label{spin EOM 1}
    \dot{\sigma}^{AB} = u^B u_C \dot{\sigma}^{AC} - u^A u_C
    \dot{\sigma}^{BC}.
\end{equation}
Here, $u^A$ is the 5-velocity, and an overdot indicates
$D/d\lambda = u^A \nabla_A$.  The equation of motion for the
5-velocity is
\begin{equation}\label{spin EOM 2}
    u^A \nabla_A u^B = A^B, \quad u^A u_A = 1,
\end{equation}
where $A^B$ is the 5D acceleration induced by non-gravitational
forces.  We have followed Schiff and neglected the coupling of
the Riemann tensor to $u^A$ as is appropriate for a point
gyroscope.  We will apply equations (\ref{spin EOM 1}) and
(\ref{spin EOM 2}) to 5D pointlike gyroscopes that are freely
falling, as described in Section \ref{sec:affine}, and gyroscopes
that are subject to a centripetal confining force, as described
in Section \ref{sec:confinement}.  In the latter case, we must
assume that the confining force acts at the center of mass to
satisfy the torque-free requirement. Essentially, we need to
neglect the ``tidal'' variation in the confining force over the
body, which is reasonable for a body of extremely small size.
With these assumptions in mind, we can apply the above spin
equations of motion to a pointlike gyroscope in 5D.

However, our analogy with familiar spacetime physics must end
here because we cannot generalize the 4D procedure of mapping the
spin tensor $\sigma^{AB}$ onto a unique spin vector $\sigma^A$ to
the 5D case. To see why this is so, we note that contraction of
equation (\ref{spin EOM 1}) with $u^A$ reveals that four of the
ten equations for $\dot{\sigma}^{AB}$ are redundant.  Hence the
system of equations (\ref{spin EOM 1}) is underdetermined; that
is, we need to impose some sort of subsidiary condition on
$\sigma^{AB}$.  As in 4D, we can choose the spin tensor to be
orthogonal to the 5-velocity
\begin{equation}
    \sigma^{AB} u_A = 0.
\end{equation}
This reduces the number of degrees of freedom in $\sigma^{AB}$ to
six.  The same requirement in 4D implies that the spin tensor has
three independent components that can be uniquely mapped onto a
4-vector orthogonal to the 4-velocity.  But in 5D, a 5-vector
orthogonal to the 5-velocity has four components, which is not
enough to describe $\sigma^{AB}$.  But, the six degrees of
freedom do correspond to the number of independent components of
an antisymmetric 4-dimensional matrix.  This motivates us to
decompose $\sigma^{AB}$ into a basis $\{ \sigma_a^A \}$ that
spans the space orthogonal to $u^A$:
\begin{equation}\label{spin tensor}
    \sigma^{AB} = \sigma^{ab} \sigma_a^A \sigma_b^B, \quad 0 =
    u_A \sigma^A_a, \quad \sigma^{ab} = -\sigma^{ba}.
\end{equation}
We now demand that the $\{ \sigma_a^A \}$ basis is chosen in a
manner that ensures that the $4 \times 4$ $\sigma^{ab}$ matrix
has constant entries:
\begin{equation}
    \nabla_A \sigma^{ab} = 0.
\end{equation}
Note that $\sigma^{ab}$ behaves like a 5D scalar quantity.
Substitution of our assumed form of $\sigma^{AB}$ into equation
(\ref{spin EOM 1}) and contracting with $\sigma^A_c$ yields,
after some algebra
\begin{equation}
    0 = \sigma^{ab} \left\{ ( \sigma_a \cdot \sigma_c )
    [ \dot{\sigma}_b + ( \dot{u} \cdot \sigma_b ) u ] + (
    \dot{\sigma}_a \cdot \sigma_c ) \sigma_b \right\},
\end{equation}
where we have suppressed the 5D indices for clarity.  This can be
solved in a manner independent of our choice of $\sigma^{ab}$ if
the basis vectors satisfy
\begin{equation}\label{Fermi-Walker}
    u^B \nabla_B \sigma^A_a = - (A_B \sigma^B_a) u^A,
\end{equation}
where we have made use of (\ref{spin EOM 2}).  This is the
equation of 5D Fermi-Walker (FW) transport of $\sigma^A_a$ along
the integral curves of $u^A$ subject to the condition $0 = u_A
\sigma^A_a$.  Therefore, we have demonstrated that the spin
tensor of a pointlike gyroscope in 5D can be expressed in the
form given in equation (\ref{spin tensor}), where $\sigma^{ab}$
is an arbitrary $4 \times 4$ antisymmetric tensor with constant
entries, provided that the $ \{ \sigma^A_a \}$ basis is FW
transported along the gyroscope's trajectory.  The six degrees of
freedom of $\sigma^{AB}$ are carried by the $\sigma^{ab}$ matrix.

We note that the spin vector of a particle in 4D is governed by
an equation identical to (\ref{Fermi-Walker}).  Therefore, the
problem of determining the evolution of the spin tensor of a
pointlike gyroscope in 4D and 5D is operationally identical;
i.e.\ one needs to solve the FW transport equation. However, the
relation between the solution(s) of that equation and the full
spin tensor is different.  It is interesting to note that the
method outlined here will work in any dimension, including 3+1,
while the procedure of identifying spin angular momentum with a
single vector is peculiar to the case of three spatial
dimensions.  Regardless, we are now faced with the prospect of
solving the FW equation in 5D.  This is the subject of the next
section.

\subsection{Splitting of the 5D Fermi-Walker transport equation}
\label{sec:fermi}

In this section, we will attempt to perform a $(3+1)+1$ splitting
of the equation of FW transport for the spin-basis vectors $\{
\sigma^A_a \}$ similar to the splitting of the geodesic equation
performed in Section \ref{sec:affine}.  The relevant formulae are
given by equations (\ref{spin EOM 2}) and (\ref{Fermi-Walker}).
For brevity, we will omit the Latin index on the spin-basis
vector $\sigma_a^A$.  We will consider the cases of free and
constrained motion in 5D by setting
\begin{equation}
    A^B =
    \begin{cases}
        0, & \text{for 5D geodesic motion,} \\
        K_{\alpha\beta} u^\alpha u^\beta n^B, & \text{for 5D confined
        paths.}
    \end{cases}
\end{equation}
As mentioned above, we assume that the spin basis is orthogonal
to $u^A$.  Since $\sigma^A$ is FW transported along the gyroscope
trajectory, its magnitude is constant and can be set to $-1$.
Hence we also have
\begin{equation}\label{5D spin constants}
    u^A \sigma_A = 0, \quad \sigma^A \sigma_A = -1.
\end{equation}
We also define
\begin{equation}\label{spin definitions}
    \sigma^\alpha = e^\alpha_A \sigma^A, \quad \varSigma =
    \sigma^A n_A.
\end{equation}

\paragraph*{Case 1: $A^B = 0$.}

The gyroscope's center of mass motion is described by equations
(\ref{split equations}).  The mechanics of the decomposition of
equations (\ref{Fermi-Walker}) and (\ref{5D spin constants}) is
similar to the calculations of Section \ref{sec:affine}, so we
will omit the details and present the final results.  We get
\begin{subequations}\label{split spin equations}
\begin{eqnarray}
    u^\alpha \nabla_\alpha \sigma^\beta & = & \varSigma
    K^{\alpha\beta} u_\alpha + \xi e^\beta_B n^A \nabla_A
    \sigma^B, \\ \label{spin extra part} \dot{ \varSigma }  & = &
    K_{\alpha\beta} u^\alpha \sigma^\beta - \xi
    n^A \sigma^B \nabla_A n_B, \\ 1 & = & \varSigma^2 -
    h_{\alpha\beta} \sigma^\alpha \sigma^\beta.
\end{eqnarray}
\end{subequations}
These formulae are analogous to the three equations (\ref{split
equations}) used to describe the behaviour of $u^\alpha$ and
$\xi$.  The fact that $\sigma^\alpha$ does not satisfy the 4D FW
transport equation means that there will appear to be an
anomalous torque acting on the 4D spin tensor.  This torque
prevents the norm of the $\sigma^\alpha$ 4-vector from being a
constant of the motion, despite the fact that norm of $\sigma^A$
is conserved.  This result causes us to ask if $u^\alpha
\sigma_\alpha$ is a conserved quantity, like $u^A \sigma_A$ is.
We note that $u^A \sigma_A = 0$ implies that $u^\alpha
\sigma_\alpha = \xi \varSigma$. Differentiating this scalar
relation with respect to $\lambda$, we get
\begin{equation}
    d(u^\alpha \sigma_\alpha)/d\lambda = \xi \dot{ \varSigma }  +
    \varSigma \dot{\xi}.
\end{equation}
We substitute in expressions for $\dot{\xi}$ and $\dot{ \varSigma
} $ from equations (\ref{extra part}) and (\ref{spin extra part})
and simplify to get
\begin{equation}\label{angle evolution}
    \frac{d}{d\lambda} (u^\alpha \sigma_\alpha) = (\xi \sigma^B +
    \varSigma u^B ) u^A \nabla_A n_B.
\end{equation}
In obtaining this equation, we have made use of the identity
\begin{equation}
    u^A \nabla_A n_B = K_{\alpha\beta} u^\alpha e^\beta_B - \xi
    n^A \nabla_A n_B,
\end{equation}
which can be proved by expanding $\nabla_A n_B$ in the basis
vectors $(e^A_\alpha,n^A)$.  Equation (\ref{angle evolution})
demonstrates that $u^\alpha \sigma_\alpha$ is not a constant of
the motion.

\paragraph*{Case 2: $A^B = K_{\alpha\beta} u^\alpha u^\beta
n^B$.}

In this eventuality, we take the 5D trajectory to be described by
equations (\ref{confined equations}).  The splitting of equations
(\ref{Fermi-Walker}) and (\ref{5D spin constants}) takes the form
\begin{subequations}\label{confined spin}
\begin{eqnarray}\label{confined spin 1}
    \tau^\beta & = & u^\alpha \nabla_\alpha \sigma^\beta, \\
    \label{confined spin extra part}
    \dot{ \varSigma }  & = & K_{\alpha\beta} u^\alpha \sigma^\beta,
    \\ \label{confined spin 3} 1 & = & \varSigma^2 -
    h_{\alpha\beta} \sigma^\alpha \sigma^\beta,
\end{eqnarray}
\end{subequations}
where we have defined the anomalous torque by
\begin{equation}\label{tau}
    \tau^\beta = \varSigma ( K^{\alpha\beta} u_\alpha - K^{\mu\nu}
    u_\mu u_\nu u^\beta).
\end{equation}
This anomalous torque satisfies
\begin{equation}\label{confined spin scalars}
    0 = \tau^\beta u_\beta, \quad \varSigma \dot{ \varSigma }  =
    \tau^\beta \sigma_\beta.
\end{equation}
The lefthand equation implies
\begin{equation}\label{conserved angle}
    0 = h_{\alpha\beta} u^\alpha \sigma^\beta;
\end{equation}
i.e.\ the angle between $u^\alpha$ and $\sigma^\alpha$ is
conserved.  As mentioned above, this not true for the case of 5D
geodesic motion.  The righthand member of (\ref{confined spin
scalars}) is consistent with (\ref{confined spin 3}); i.e.\ the
magnitude of $\sigma^\alpha$ is not conserved.

We have hence derived equations (\ref{split spin equations} and
\ref{confined spin}) governing the behaviour of the $\{
\sigma_a^A \}$ spin basis in terms of a $(3+1)+1$ splitting of
the 5D manifold.  However, 4D observers will not observe these
vectors directly, they will rather see the projection of the spin
tensor $\sigma^{AB}$ onto $\Sigma_\ell$.  So, to make contact
with physics, we must consider
\begin{equation}\label{4D spin tensor}
    \sigma^{\alpha\beta} = e^\alpha_A e^\beta_B \sigma^{AB} =
    \sigma^{ab} \sigma^\alpha_a \sigma^\beta_b,
\end{equation}
where we have made use of the decomposition of $\sigma^{AB}$
given by equation (\ref{spin tensor}) and equation (\ref{spin
definitions}) and re-introduced the spin basis indices.  We can
now ask various physical questions, for example, is the magnitude
of the 4D spin tensor conserved?  We can write
\begin{equation}
    \sigma_{\alpha\beta} \sigma^{\alpha\beta} = h_{AB} h_{CD}
    \sigma^{AC} \sigma^{BD}.
\end{equation}
Expanding $h_{AB}$ and simplifying yields
\begin{equation}\label{4D spin variation}
    \sigma_{\alpha\beta} \sigma^{\alpha\beta} = \sigma_{ab}
    \sigma^{ab} + 2 \sigma^{ab} \sigma_{cb} \varSigma^c
    \varSigma_a,
\end{equation}
where we have defined
\begin{equation}
    \varSigma_a = n_A \sigma^A_a
\end{equation}
and used the metric of the spin basis
\begin{equation}
    q_{ab} = g_{AB} \sigma^A_a \sigma^B_b
\end{equation}
to raise and lower spin indices.  It is easy to demonstrate that
$\nabla_A q_{ab} = 0$ from equation (\ref{Fermi-Walker}), so
$\sigma_{ab} \sigma^{ab}$ is a constant.  Therefore, the 4D spin
$\sigma_{\alpha\beta} \sigma^{\alpha\beta}$ will not be conserved
if $\varSigma_a$ varies along the path, as is the case for both
freely falling and constrained trajectories (equations \ref{spin
extra part} and \ref{confined spin extra part}).

Clearly, the behaviour of $\sigma^{\alpha\beta}$ in the general
case is a subject that deserves in-depth study, but we will defer
such discussions to future work.  We will instead give a specific
example of how the magnitude of the 4D spin of a gyroscope will
vary when that gyroscope is confined to a $3+1$ dimensional
hypersurface in a $(3+1) + 1$ dimensional manifold.  This example
is the subject Section \ref{sec:example}.

\subsection{Gyroscopes in the brane world}\label{sec:brane}

As in Section \ref{sec:confinement}, the observant reader will
have again noticed that we will have trouble applying our results
to the thin brane world scenario.  However, our predicament is
more dire in this situation, because the equations (\ref{confined
spin}) governing the evolution of spin basis vectors confined to
a single $\Sigma_\ell$ hypersurface make explicit reference to
the extrinsic curvature of the that hypersurface.  This is a
problem because, as seen in equation (\ref{K jump}), the
extrinsic curvature of an infinitely thin brane is ill-defined.
At best, our formulae can be used to describe gyroscopes
traveling on the $\gamma_\pm$ bulk trajectories discussed in
Section \ref{sec:confinement}, which are infinitesimally above or
below the brane.

Before abandoning the thin brane world entirely, we can try to
understand the behaviour of a spin-basis vector in the vicinity
of the brane.  As in Section \ref{sec:confinement}, we place the
brane at the position of the $\Sigma_0$ hypersurface and consider
the neighboring $\Sigma_\pm$ hypersurfaces.  The curves
$\gamma_0$ and $\gamma_\pm$ are geodesics on the respective
hypersurfaces, and we have previously seen that $\gamma_\pm
\rightarrow \gamma_0$ as $\Sigma_\pm \rightarrow \Sigma_0$.  Now,
consider the spin basis vectors $\sigma_\pm^A$, which are 5D
vectors FW transported along $\gamma_\pm$, and hence satisfy
equations (\ref{Fermi-Walker}) and (\ref{confined spin}). As
mentioned before, the 5D acceleration of the $\gamma_\pm$ curves
differs by a sign, which implies that $\sigma_\pm^A$ will not
have the same equation of motion.  Hence, we will in general have
that $\sigma_+^A \nrightarrow \sigma_-^A$ as $\Sigma_\pm
\rightarrow \Sigma_0$, irrespective of initial conditions.  To
state this in a different way, imagine that $\gamma$ is a 5D
congruence of curves which are everywhere tangent to geodesics of
$\Sigma_\ell$ hypersurfaces and that $\sigma^A$ is a 5D vector
field that is everywhere FW transported along $\gamma$. Then, we
have seen that $\gamma$ can be chosen to be smooth across the
brane but that $\sigma^A$ is in general \emph{discontinuous}
across the $\Sigma_0$ hypersurface.

\begin{figure}
\includegraphics{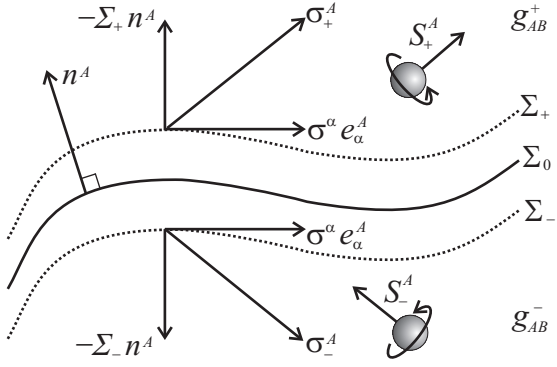}
\caption{The spin basis vector field $\sigma^A$ in the
neighborhood of the brane $\Sigma_0$. $\sigma^A_\pm$ are mirror
images of one another, which is what one might expect of an
ordinary vector under the $\mathbb{Z}_2$ symmetry.  For
reference, we show how an axial spin vector $S^A$ would transform
under $\mathbb{Z}_2$, with its tangential components reversed.
}\label{fig:mirror}
\end{figure}

What is the nature of this discontinuity?  It is easily seen that
while equations (\ref{confined spin}) are not invariant under
$K_{\alpha\beta} \rightarrow -K_{\alpha\beta}$, they are
invariant under $(K_{\alpha\beta} \rightarrow
-K_{\alpha\beta},\sigma^\alpha \rightarrow -\sigma^\alpha)$ and
$(K_{\alpha\beta} \rightarrow -K_{\alpha\beta},\varSigma
\rightarrow -\varSigma)$.  This raises two possibilities, either
the normal component of $\sigma^A$ is continuous and the
tangential component is discontinuous across the brane or
\emph{vice versa}.  The former situation is akin to the behaviour
of the magnetic field in the presence of a surface current, while
the latter case is like discontinuity of the electric field
around a surface charge distribution.  In both cases, the
discontinuous component is reversed as the brane is traversed. To
define the continuation of the $\sigma^A$ from, say, the $+$ side
of the brane to the $-$ side, we need to choose which component
is continuous and which is not.  There is no mathematical reason
to prefer one choice over the other, but an intuitive choice
comes from the $\mathbb{Z}_2$ symmetry around the brane.  This
symmetry implies that we can think of $\Sigma_0$ as a mirror.
Hence, if we choose to have the tangential component of
$\sigma^A$ reversed on either side of the brane we have
essentially elected to have $\sigma^A$ transform as an axial
vector (also known as a psuedovector) under reflections.  The
opposite choice, namely that the normal component of $\sigma^A$
is reversed as one crosses the brane, implies that $\sigma^A$
transforms as an ordinary vector under reflections.  The former
situation is what we would expect if $\sigma^A$ were a
traditional spin vector.  But $\sigma^A$ is not a 5D spin vector,
it is simply a member of a basis and should hence transform as an
ordinary vector. Therefore, we choose to have the tangential
components of $\sigma^A$ continuous across the brane. Our choice
for the continuation of $\sigma^A$ across $\Sigma_0$ is shown in
figure \ref{fig:mirror}, along with the alternate scenario for a
hypothetical 5D spin vector $S^A$.

This continuation of $\sigma^A$ across the brane can be viewed as
a way out of our dilemma because the observationally accessible
part of $\sigma^A$ has a well defined limiting value as $\ell
\rightarrow 0$.  In this case, the dynamics of $\sigma^\alpha$
and $\varSigma$ can be worked out using equations (\ref{confined
spin}) on either the $+$ or $-$ side with the assurance that the
answer for $\sigma^\alpha$ will be the same.  Whether or not this
mathematical trick has any physical relevance is an open
question; the skeptical reader may conclude that the
discontinuity in the geometry precludes sensible descriptions of
5D spin tensors on the brane, which could be viewed as an
indictment of the thin-brane picture.  At any rate, our formalism
\emph{can} be freely applied to smooth manifolds, which include
thick-brane solutions.  In the next section, we will look at a
specific example of gyroscope motion from 5D non-compactified
Kaluza-Klein theory.

\subsection{Variation of 4D spin in a cosmological setting}
\label{sec:example}

We now turn our attention to a specific metric which has been
used to embed standard $k=0$
Friedmann-Lema{\^\i}tre-Robertson-Walker (FLRW) cosmologies in a
5D flat space.  The line element, which was first given by Ponce
de Leon \cite{Pon88}, is
\begin{equation}\label{Ponce}
    d\mathcal{S}^2 = \ell^2 \, dt^2 - a^2(t,\ell) d\roarrow{x}
    \cdot d\roarrow{x} - \frac{\alpha^2 t^2}{(1-\alpha)^{2}}
    \, d\ell^2,
\end{equation}
where $\roarrow{x} = (x,y,z)$, $\alpha$ is a parameter, and
\begin{equation}
    a(t,\ell) = t^{1/\alpha} \ell^{1/(1-\alpha)}.
\end{equation}
This metric is a solution of $R_{ABCD} = 0$; i.e.\ it is 5D
Minkowski space written in complicated coordinates.  The $\ell =$
constant hypersurfaces of this metric share the same geometry as
a $k = 0$ FLRW solution with a scale factor equal to $a(t,\ell)$,
which in turn corresponds to matter with an equation of state $p
= (2 \alpha / 3 - 1) \rho$.  The way in which the 4D big bang is
embedded by metrics of the form (\ref{Ponce}) has been discussed
in detail elsewhere \cite{Wes01,Sea01b}.

In this section, we will consider a pointlike gyroscope confined
to move on one of the $\Sigma_\ell$ hypersurfaces by a
non-gravitational centripetal force as discussed in Section
\ref{sec:confinement}.  Our goal will be to solve equations
(\ref{confined spin}) for the orbits of the spin basis vectors
$\{ \sigma^\alpha_a,\varSigma_a \}$.  We make the simplest
possible choice for coordinates on $\Sigma_\ell$, namely
$y^\alpha = (t,x,y,z)$ so that $e^A_\alpha = \delta^A_\alpha$.
Let us introduce a set of basis vectors on $\Sigma_\ell$:
\begin{subequations}
\begin{eqnarray}\label{lambda 0 def}
    \lambda^\alpha_{(0)} & = & u^\alpha = [\ell^{-1}f(t,\ell),
    \beta a^{-2}(t,\ell),0,0], \\
    \lambda^\alpha_{(1)} & = & \bar{u}^\alpha = a^{-1}(t,\ell)
    [\beta \ell^{-1},f(t,\ell),0,0], \\
    \lambda^\alpha_{(2)} & = & \hat{y}^\alpha = a^{-1}(t,\ell)
    [0,0,1,0], \\ \lambda^\alpha_{(3)} & = &
    \hat{z}^\alpha = a^{-1}(t,\ell) [0,0,0,1].
\end{eqnarray}
\end{subequations}
Here, $\beta$ is a parameter and the function $f(t,\ell)$ is
given by
\begin{equation}
    f(t,\ell) = \sqrt{ 1 + \beta^2 a^{-2}(t,\ell) }.
\end{equation}
It is not difficult to verify that this basis is orthonormal:
\begin{equation}
    \eta_{(\mu)(\nu)} = h_{\alpha\beta} \lambda^\alpha_{(\mu)}
    \lambda^\beta_{(\nu)},
\end{equation}
where $\eta_{(\mu)(\nu)} = \mathrm{diag}(1,-1,-1,-1)$. Also, one
can verify that the basis is parallelly-propagated along the
integral curves of $u^\alpha$:
\begin{equation}
    u^\alpha \nabla_\alpha \lambda^\beta_{(\mu)} = 0.
\end{equation}
Hence, $u^\alpha$ is tangent to geodesics on $\Sigma_\ell$ and
the other members of $\{ \lambda^\alpha_{(\mu)} \}$ are 4D FW
transported along those geodesics.  These geodesics represent a
particle moving in the $x$-direction with a proper speed of
$\beta/a^2(t,\ell)$.  As demonstrated in Section
{\ref{sec:confinement}, when test particles are confined to a
given $\Sigma_\ell$ hypersurface they will travel on geodesics of
that hypersurface.  Therefore, we can take our gyroscope to be
traveling on one of the $u^\alpha$ integral curves.  Also notice
that if the spin basis were 4D FW transported along the integral
curves of $u^\alpha$, the projections of $\sigma_a^\alpha$ onto
$\{ \lambda^\alpha_{(\mu)} \}$ basis would be constant. We shall
see that this is not the case for 5D FW transport.

Having specified the form of the trajectory, we turn our
attention to equations (\ref{confined spin}).  By calculating the
extrinsic curvature of the $\Sigma_\ell$ hypersurfaces and
substituting in the expression (\ref{lambda 0 def}) for
$u^\alpha$, we can determine the anomalous torque defined by
equation (\ref{tau}):
\begin{equation}\label{cosmic tau}
    \tau^\alpha = \frac{\beta \varSigma}{t\ell} \left[
    \frac{f(t,\ell)}{a(t,\ell)} \right] \bar{u}^\alpha.
\end{equation}
Interestingly enough, the anomalous torque vanishes if the
gyroscope is comoving with $\beta = 0$.  This is demanded by
isotropy; a nonzero $\tau^\alpha$ for comoving paths would pick
out a preferred spatial direction.

Continuing, we suppress the Latin index on $\sigma^\alpha_a$ and
$\varSigma_a$.  Now, equation (\ref{conserved angle}) gives that
$\sigma^\alpha$ is orthogonal to $u^\alpha$.  We can therefore
expand any spin basis vector as follows
\begin{equation}
    \sigma^\alpha = \sigma \hat{\sigma}^\alpha, \quad -1 =
    \hat{\sigma}_\alpha \hat{\sigma}^\alpha, \quad 0 = u_\alpha
    \hat{\sigma}^\alpha,
\end{equation}
with
\begin{equation}
    \hat{\sigma}^\alpha = \bar{u}^\alpha \cos\theta +
    \sin\theta ( \hat{y}^\alpha \cos\phi + \hat{z}^\alpha \sin\phi ).
\end{equation}
Here, $(\sigma,\theta,\phi)$ are considered to be functions of
$t$ and $\ell$ and can be thought of as the spherical polar
components of $\sigma^\alpha$.  Equation (\ref{confined spin 3})
gives
\begin{equation}
    1 = \sigma^2 + \varSigma^2,
\end{equation}
which motivates the ansatz
\begin{equation}\label{gamma def}
    \sigma = \sin\gamma(t,\ell), \quad \varSigma = \cos\gamma(t,\ell).
\end{equation}
Figure \ref{fig:spin} depicts the decomposition of $\sigma^A$ in
the $\{\lambda^\alpha_{(i)},n^A\}$ basis.  (Recall that since
$u^A \sigma_A = 0$, $\sigma^\alpha$ will have no projection on
$\lambda^\alpha_{(0)} = u^\alpha$.)
\begin{figure}
\includegraphics[width=2.4in]{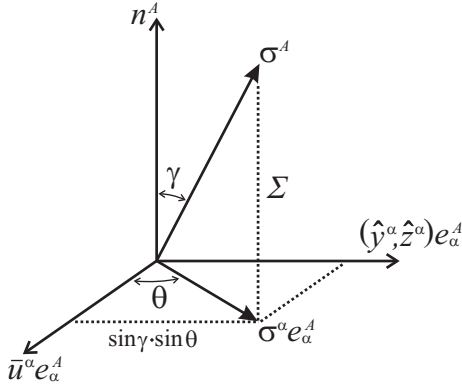}
\caption{The decomposition of $\sigma^A$ in the
$\{\lambda^\alpha_{(i)},n^A\}$ basis.}\label{fig:spin}
\end{figure}

Substituting equations (\ref{cosmic tau}) -- (\ref{gamma def})
into equation (\ref{confined spin 1}) and taking scalar products
with each of $\{ \lambda^\alpha_{(\mu)} \}$ basis vectors yields
an integrable set of first order differential equations for
$(\gamma,\theta,\phi)$.  We omit the details and quote the
results:
\begin{subequations}
\begin{eqnarray}\label{spin sol'n 1}
    \cos\gamma & = & \cos \varphi_1 \cos[\varphi_2 + \alpha \beta
    a^{-1}(t,\ell)],
    \\ \label{spin sol'n 2} \sin \varphi_1 & = & \sin\gamma \sin\theta,
    \\ \label{spin sol'n 3} \phi & = & \varphi_3,
\end{eqnarray}
\end{subequations}
where we require
\begin{equation}
    -\tfrac{\pi}{2} \le \varphi_1 \le \tfrac{\pi}{2},
    \quad \sin\gamma \ne 0.
\end{equation}
Here, the angles $\{ \varphi_i \}$ are constants of integration.
Equation (\ref{spin sol'n 1}) governs the evolution of
$\sigma^\alpha \sigma_\alpha$ and $\varSigma$, while equations
(\ref{spin sol'n 2}) and (\ref{spin sol'n 3}) state that the
projection of $\sigma^\alpha$ onto the plane spanned by
$\hat{y}^\alpha$ and $\hat{z}^\alpha$ is of constant magnitude.
In the late epoch limit we have that $a(t,\ell) \rightarrow
\infty$, which implies that $\cos\gamma$ and $\sin\theta$
approach constant values.  In other words, the spin basis vectors
become static for late times.  As mentioned above, they are also
static for comoving gyros with $\beta = 0$.  For early times, the
variation of the $\gamma$ and $\theta$ angles implies that the
spin basis vector precess with respect to a 4D non-rotating
frame.

To make contact with 4D physics, we must now specify a set of
four linearly independent spin basis vectors by choosing four
different sets of the constant angles $\{ \varphi_i \}$.  We can
then construct a 4D spin tensor from equation (\ref{4D spin
tensor}), with $\sigma^{ab}$ arbitrary.  We will not do that
explicitly here, we rather content ourselves with the observation
that equations (\ref{4D spin tensor}), (\ref{4D spin variation}),
(\ref{gamma def}) and (\ref{spin sol'n 1}) imply that the
magnitude of the 4D spin
$\sigma^{\alpha\beta}\sigma_{\alpha\beta}$ is not conserved.  In
fact, it is not difficult to show that in the $a \rightarrow
\infty$ limit the derivative of the spin magnitude obeys
\begin{equation}
    \left| \frac{d}{da} \sigma^{\alpha\beta}\sigma_{\alpha\beta}
    \right| \propto a^{-2}.
\end{equation}
Because this variation takes place on cosmic timescales, it is
not likely to be observed by an experiment like Gravity Probe B.
However, the example in this section was intended to be
illustrative of the method rather than an experimental
suggestion.  Application of the formalism to other higher
dimensional scenarios may lead to experimentally or
observationally testable effects.  We hope to report on such
matters in the future.

\section{Summary and Conclusions}\label{sec:conclusions}

In this paper, we have used the $(3+1)+1$ dimensional foliation
of a non-compact 5D manifold described in Section
\ref{sec:geometry} to analyze various aspects of test particle
and pointlike gyroscope motion in higher dimensions.

In Section \ref{sec:affine}, we split the 5D
affinely-parameterized geodesic equation into a 4D equation of
motion (\ref{4D part}), an equation governing the motion
perpendicular to $\Sigma_\ell$ (\ref{extra part}), and an
equation describing the evolution of the norm of the 4-velocity
(\ref{split norm}).  We also demonstrated that these three
equations were not independent.  In Section \ref{sec:proper}, we
described how equations (\ref{split equations}) behave under a
general change of parameter (equations \ref{transformed split
equations}).  We then gave their form in the 4D proper time
parameterization (equations \ref{s acceleration} and \ref{s
extra}).  In the latter case, we saw that the 4-velocity was
properly orthogonal to the 4-acceleration.  In Section
\ref{sec:differences}, we showed that the projected 4-velocity
$u^\alpha = e^\alpha_A u^A$ does not equal $dy^\alpha /
d\lambda$, but rather corresponds to the velocity in canonical
coordinates.  We also saw that the fifth force $f^\alpha$ defined
by equation (\ref{f definition}) is not equal to the
4-acceleration and does not transform as a 4-vector under
$y^\alpha \rightarrow \tilde{y}^\alpha(y^\beta)$.

In Section \ref{sec:confinement}, we derived the form of the
force required to confine a particle to a single $\Sigma_\ell$
hypersurface and showed that it reduced to the ordinary
centripetal force in Minkowski space.  We also demonstrated that
particles travel on geodesics of $\Sigma_\ell$ under these
conditions.

In Section \ref{sec:spin}, we showed how the problem of
determining the 5D orbit of a pointlike gyroscope can be reduced
to the solution of the Fermi-Walker transport equation, just as
in 4D, but relation to the spin tensor is different than in the
spacetime case.  In Section \ref{sec:fermi}, we performed a
$(3+1)+1$ split of the 5D FW equation for the case of freely
falling gyroscopes (equation \ref{split spin equations}) and
confined gyroscopes (equation \ref{confined spin}).  We
demonstrated that in both cases, the magnitude of the 4D spin is
not conserved due to the existence of an anomalous torque.  In
Section \ref{sec:brane}, we discussed how our results should be
interpreted in the thin brane world scenario.  In Section
\ref{sec:example}, we applied our formulae to a specific 5D
cosmological example and derived how the 4D spin of a gyroscope
varies when confined to a $\Sigma_\ell$ hypersurface.

In conclusion, we mention a few possible directions for future
work. Equations (\ref{split equations}) can be used to study the
motion of observers in the thick brane world and non-compact
Kaluza-Klein theories, apparent violations of 4D causality due to
the existence of 5D ``short-cuts'', and the effect of 5D dynamics
on astrophysical systems.  Equation (\ref{extra part}) can be
used to study the issue of whether a given 3-brane attracts or
repels test particles.  On the theoretical side, an interesting
exercise involves determining how the extra acceleration
$a^\alpha(u)$ encodes the electromagnetic force previously
observed in the fifth force $f^\alpha(u)$ derived from the 5D
geodesic equation.  The discrepancy between the 5D affine
parameter and the 4D proper time seen in Section \ref{sec:proper}
raises the question of which one is the correct ``clock'' to use,
which certainly merits close attention. Our formalism concerning
pointlike gyroscopes should be applied to 5D static and
spherically-symmetric metrics in order to make predictions
testable by Gravity Probe B.  The issue of the cosmological
variation of spin can be applied to the evolution of the angular
momentum of galaxies, pulsars and high-energy primordial objects.
These ideas do not comprise an exhaustive list of potential
avenues of exploration, which underlines the generality and wide
applicability of formulae derived in this paper.

\begin{acknowledgments}
We would like to thank P.\ S.\ Wesson, J.\ Ponce de Leon and D.\
Bruni for useful conversations and NSERC for financial support.
\end{acknowledgments}

\bibliography{text}

\end{document}